\documentclass{PoS}

\newcommand{\nasyr}{$\mu$as~yr$^{-1}$}

\title{Towards a kinematic model of the Local Group}

\ShortTitle{Towards a kinematic model of the Local Group}

\author{\speaker{Andreas Brunthaler}%
         \thanks{This research was supported by the DFG Priority Programme 1177.}\\
        Max-Planck Institut f\"ur Radioastronomie\\
        E-mail: \email{brunthal@mpifr-bonn.mpg.de}}
\author{Mark J. Reid\\
        Harvard-Smithsonian Center for Astrophysics\\
        E-mail: \email{reid@cfa.harvard.edu}}
\author{Heino Falcke\\
        ASTRON\\
        E-mail: \email{falcke@astron.nl}}
\author{Christian Henkel\\
        Max-Planck Institut f\"ur Radioastronomie\\
        E-mail: \email{p220hen@mpifr-bonn.mpg.de}} 
\author{Karl M. Menten\\
        Max-Planck Institut f\"ur Radioastronomie\\
        E-mail: \email{kmenten@mpifr-bonn.mpg.de}}

\abstract{
  Measuring the proper motions and geometric distances of galaxies
  within the Local Group is very important for our understanding of the
  history, present state and future of the Local Group. Currently,
  proper motion measurements using optical methods are limited only to
  the closest companions of the Milky Way. However, Very Long Baseline
  Interferometry (VLBI) provides the best angular resolution in
  astronomy and phase-referencing techniques yield astrometric
  accuracies of $\approx$ 10 micro-arcseconds. This makes a measurement
  of proper motions and angular rotation rates of galaxies out to a
  distance of $\sim$ 1 Mpc feasible.

  This article presents results of VLBI observations of regions of
  H$_2$O maser activity in the Local Group galaxies M33 and IC\,10. 
  Two masing regions in M33 are on opposite sides of the galaxy. This 
  allows a comparison of the angular rotation rate (as measured by the VLBI
  observations) with the known inclination and rotation speed of the
  HI gas disk. This gives a geometric distance of 730 $\pm$ 100 $\pm$
  135 kpc. The first error indicates the statistical error from the
  proper motion measurements while the second error is the systematic
  error from the rotation model. This distance is consistent, within
  the errors, with the most recent Cepheid distance to M33. Since all
  position measurements were made relative to an extragalactic
  background source, the proper motion of M33 has also been
  measured. This provides a three dimensional velocity vector of M33,
  showing that this galaxy is moving with a velocity of 190 $\pm$ 59
  km~s$^{-1}$ relative to the Milky Way. For IC\,10, we obtain a motion
  of 215 $\pm$ 42 km~s$^{-1}$ relative to the Milky Way. 
  These measurements promise a
  new handle on dynamical models for the Local Group and the mass and
  dark matter halo of Andromeda and the Milky Way.
}

\FullConference{8th European VLBI Network Symposium\\
		 September 26-29, 2006\\
		 Toru\'n, Poland}

\begin{document}

\section{Introduction}
\begin{figure}
\begin{center}
\includegraphics[width=10.0cm,clip=,angle=0]{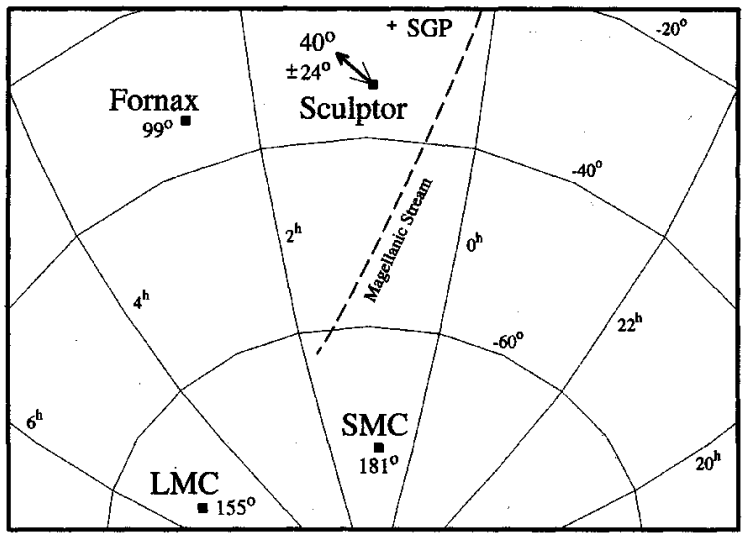}
\caption{The proper motion of the Sculptor Dwarf Spheroidal
  Galaxy. Taken from \protect\cite{6}.}
\label{sculptor}
\end{center}
\end{figure}

The nature of spiral nebulae like M33 was the topic of a famous
debate in the 1920's. While some astronomers favored a close distance
and Galactic origin, others were convinced of the extragalactic nature.
In 1923, van Maanen claimed to have measured a
large proper motion and angular rotation of M33 from photographic
plates separated by $\approx$ 12 years \cite{1}. These measurements yielded
rotational motions of $\approx$ 10--30 mas yr$^{-1}$, clearly indicating
a close distance to M33. However, Hubble discovered
Cepheids in M33, showing a large distance to M33 and confirming that
M33 is indeed an extragalactic object \cite{2}.  The expected proper motions
from the rotation of M33 are only $\approx$ 30 \nasyr, 3 orders of
magnitude smaller than the motions claimed by van Maanen. The source of
error in van Maanens motion determinations was never identified. After 
more than 80 years, the idea behind the experiment to measure the rotation and
proper motions of galaxies remains interesting for our understanding
of the dynamics and geometry of the Local Group. Hence, they are an
important science goal of future astrometric missions (e.g. Gaia~\cite{3}).

The problem when trying to derive the gravitational potential of the
Local Group is that usually only radial velocities are known and hence
statistical approaches have to be used. Kulessa and Lynden-Bell
introduced a maximum likelihood method which requires only the
line-of-sight velocities, but it is also based on some assumptions
(eccentricities, equipartition) \cite{4}.

Clearly, the most reliable way of deriving masses is using orbits,
which requires the knowledge of three-dimensional velocity vectors
obtained from measurements of proper motions.
However, measuring proper motions of members of the Local Group to
determine its mass is difficult. For the LMC a proper motion of $1.2\pm0.3$
mas~yr$^{-1}$ was obtained from comparing photographic plates over a 
time-span of 14 years \cite{5}. The Sculptor dwarf spheroidal galaxy moves 
with $0.56\pm0.25$ mas~yr$^{-1}$ obtained from plates spanning 50
years in time (Fig~\ref{sculptor}, \cite{6}). It was shown that the
inclusion of these marginal proper motions can already significantly
improve the estimate for the mass of the Milky Way, since it reduces
the strong ambiguity caused by Leo I, which can be treated as either
bound or unbound to the Milky Way \cite{7}.

In recent years, the proper motions of a number of Galactic satellites have 
been measured using the HST \cite{8,9,10,11,12,13,14,15,16}.
These galaxies are all closer than 150 kpc
and show motions between 0.2 and a few milliarcseconds (mas) per year. More
distant galaxies, such as galaxies in the Andromeda subgroup at distances of
$\sim$ 800 kpc, have smaller angular motions, which are currently not
measurable with optical telescopes.

\section{Proper Motions with VLBI}

\begin{figure}
\begin{center}
\includegraphics[width=7cm,clip=,angle=0]{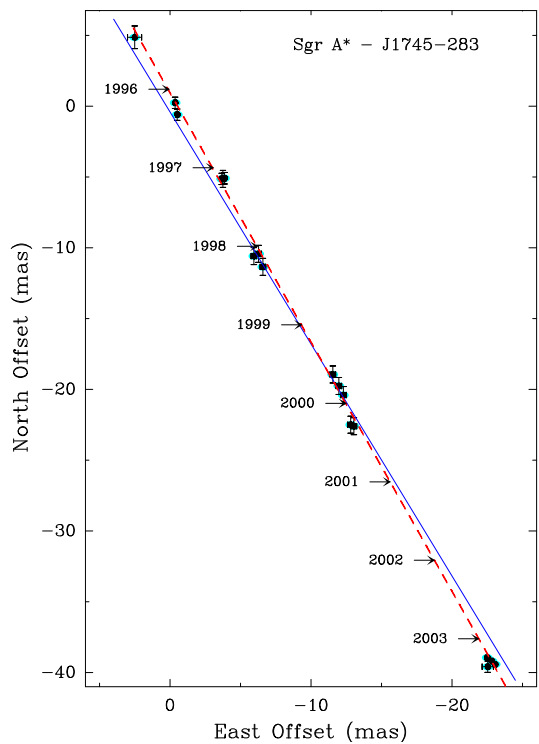}
\caption{Apparent motion of Sgr A* relative to a background quasar. The
  blue line indicates the orientation of the Galactic plane and the
  dashed line is the variance-weighted best-fit proper motion. Taken from 
 \protect\cite{18}.}
\label{sgra}
\end{center}
\end{figure}

On the other hand, the expected proper motions for galaxies within the
Local Group, ranging from 0.02 -- 1 mas yr$^{-1}$, can be seen
with Very Long Baseline Interferometry (VLBI) using the
phase-referencing technique. A good  reference point is the motion of
Sgr A* across the sky  reflecting the Sun's rotation around the
Galactic Center. This motion is well detected between epochs separated
by only one month with the
Very Long Baseline Array (VLBA)~\cite{17}. New observations increased
the time span to 8 years \cite{18}. The position of Sgr A* relative to the
background quasar J1745-283 is shown in Figure~\ref{sgra}.

The motion of Sgr A* is 6.379 $\pm$ 0.026 mas~yr$^{-1}$ and almost
entirely in the plane of the Galaxy. This gives, converted to Galactic
coordinates a proper motion of --6.379 $\pm$ 0.026  mas~yr$^{-1}$ in
Galactic longitude and --0.202 $\pm$ 0.019  mas~yr$^{-1}$ in Galactic
latitude. If one assumes a distance to the Galactic Center (R$_0$) of 
7.62 $\pm$ 0.32 kpc \cite{19}, these motions translate to a
velocity of --230 $\pm$10 km~s$^{-1}$ along the Galactic plane and --7.6
$\pm$ 0.6 km~s$^{-1}$ out of the plane of the Galaxy. This motion can
be entirely explained by a combination of a circular rotation of the
Local Standard of Rest (LSR) $\Theta_0$ and the deviation of the motion
of the Sun from the motion of the LSR.  Removing the Solar motion
relative to the LSR as measured by Hipparcos data
in~\cite{20} (5.25 $\pm$ 0.62 km~s$^{-1}$ in longitude
and --7.17 $\pm$ km~s$^{-1}$ in latitude) yields an estimate of
$\Theta_0$=225$\pm$10 km~s$^{-1}$. Here the error is dominated by the
uncertainty in the distance R$_0$ to the Galactic Center. The motion of
Sgr A* out of the plane of the Galaxy is only --0.4 $\pm$ 0.8
km~s$^{-1}$. This lower limit can be used to put tight constraints on
the mass of Sgr~A* (for details see~\cite{18}).

\begin{figure}
\begin{center}
\includegraphics[width=12cm,angle=0]{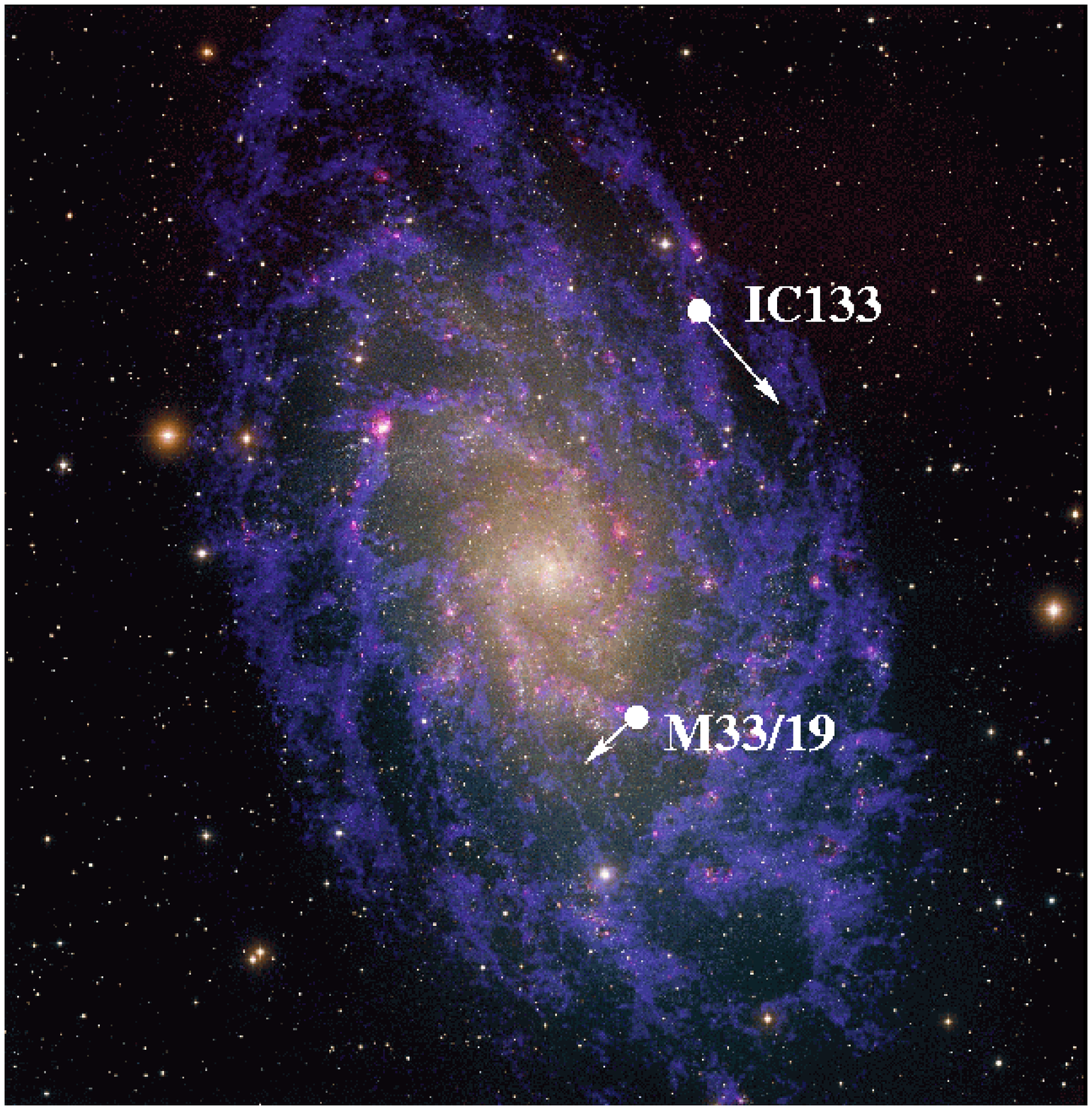}
\end{center}
\caption{Position of two regions of maser activity in M33. Also shown are
  predicted motions due to rotation of the H\,I disk (shown in blue). The 
optical image of M33 was provided by Travis Rector (NRAO/AUI/NSF and NOAO/AURA/NSF), and the H\,I image by David Thilker (NRAO/AUI/NSF) and Robert Braun 
(ASTRON). Taken from \protect\cite{22}.}
\label{m33_pred}
\end{figure}

With the accuracy obtainable with VLBI one can in principle measure
very accurate proper motions for most Local Group members within less
than a decade. The main problem so far is finding appropriate radio
sources. Useful sources would be either compact radio cores or strong
maser lines associated with star forming regions. Fortunately, in a
few galaxies bright masers are already known. 

The most suitable candidates for such a VLBI phase-referencing
experiment are the strong H$_2$O masers in IC\,10 ($\sim$ 10 Jy peak
flux in 0.5 km~s$^{-1}$ line) and IC\,133 in M33 ($\sim2$ Jy,
the first extragalactic maser discovered) \cite{21}.
The two galaxies belong to the brightest members of the Local Group
and are thought to be associated with M31. In both cases a relatively
bright phase-referencing source is known to exist within a degree.  In
addition, their galactic rotation is well known from H\,I
observations. Consequently, M33 and IC\,10 seem to be the best known
targets for attempting to measure Local Group proper motions with the
VLBA.

\section{VLBI observations of M33 and IC\,10}

\subsection{Observations and Data Reduction}

We observed two regions of H$_2$O maser activity in M33 (M33/19
and IC\,133) eight times with the NRAO Very Long Baseline Array (VLBA) 
between March 2001 and June 2005 \cite{22}. M33/19 is located in the 
south-eastern part of M33, while IC\,133 is located in the north-east of 
M33 (see Fig.~\ref{m33_pred}). 
We observed the usually brightest maser in IC\,10-SE with the VLBA 
thirteen times between February 2001 and June 2005 \cite{28}. 

The observations involved rapid switching between the phase calibrator and 
the target sources With source changes every 30 seconds, an integration time
of 22 seconds was achieved per scan.  From the second epoch on, we included 
{\it geodetic-like} observations where
we observed for 45 minutes 10--15 strong radio sources ($>$ 200 mJy) with
accurate positions ($<$ 1 mas) at different elevations to estimate an
atmospheric zenith delay error in the VLBA calibrator model (see
\cite{18,23} for a discussion). In the second and third epoch we used two 
blocks of these geodetic observations before and after the phase-referencing 
observations. From the fourth epoch on, we included a third geodetic block 
in the middle of the observation. The data were edited and calibrated using
standard techniques in the Astronomical Image Processing System (AIPS), and
zenith delay corrections were applied to improve the
accuracy of the phase-referencing. The masers in IC\,133, M33/19, and IC\,10
were imaged with standard techniques in AIPS.

\subsection{Proper Motions of M33/19 and IC\,133}

\begin{figure}
\begin{center}
\includegraphics[width=10.5cm,angle=0]{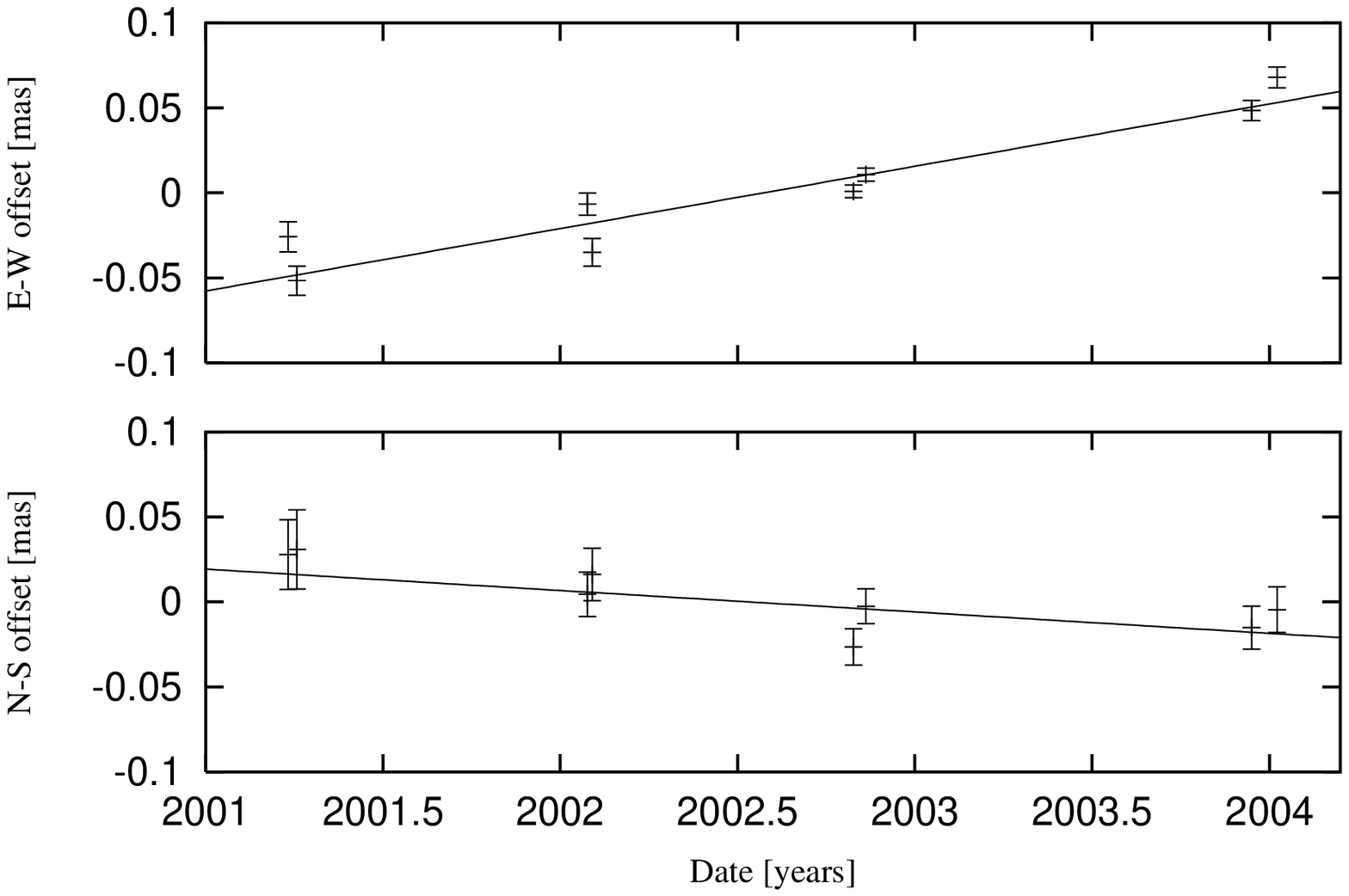}
\caption{Average  positions of the maser in M33/19 in East-West (upper) and 
North-South (lower) relative to the reference source. Taken from 
\protect\cite{22}.}
\end{center}
\begin{center}
\includegraphics[width=10.5cm,angle=0]{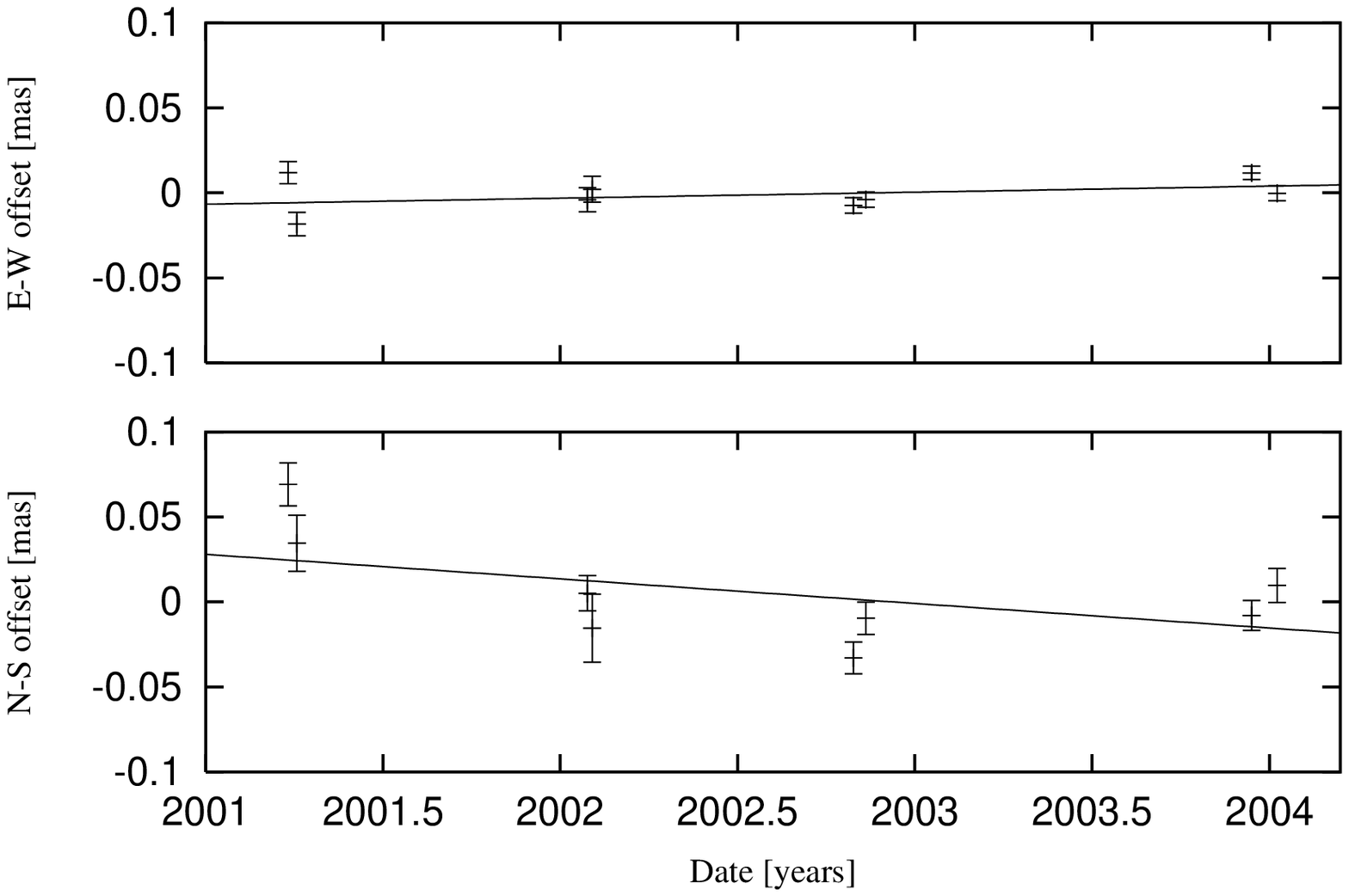}
\caption{Average  positions of the maser in IC\,133 in East-West (upper) and 
North-South (lower) relative to the reference source. Taken from 
\protect\cite{22}.}
\label{m33}
\end{center}
\end{figure}

The maser emission in M33/19 and IC\,133 is variable on timescales
of less than one year. Between the epochs, new maser features appeared while
others disappeared. However, the motions of 4 components in M33/19 and
6 components in IC\,133 could be followed over all four epochs. The
component identification was based on the positions and radial
velocities of the maser emission. Each component was usually detected
in several frequency channels. A rectilinear motion was fit to each
maser feature in each velocity channel separately. Fits with a reduced
$\chi^2$ larger than 3 were discarded as they are likely affected by
blending. Then, the variance weighted average of all motions was
calculated.  This yields an average motion of the maser components in
M33/19 of 35.5 $\pm$ 2.7 $\mu$as yr$^{-1}$ in right ascension and
$-$12.5 $\pm$ 6.3 $\mu$as yr$^{-1}$ in declination relative to the
background source J0137+312. For IC\,133 one gets an
average motion of 4.7 $\pm$ 3.2 $\mu$as yr$^{-1}$ in right ascension
and $-$14.1 $\pm$ 6.4 $\mu$as yr$^{-1}$ in declination.

\subsection{Geometric Distance of M33}

The relative motions between M33/19 and IC\,133 are independent of the
proper motion of M33 and any contribution from the motion of the
Sun. Since the rotation curve and inclination of the galaxy disk are
known one can predict the expected relative angular motion of the two
masing regions. The rotation of the H\,I gas in M33 has been measured
\cite{24} and one can calculate the expected
transverse velocities of M33/19 and IC\,133. This gives a relative
motion of 106.4 km~s$^{-1}$ in right ascension and 35 km~s$^{-1}$ in
declination between the two regions of maser activity.

The radial velocities of the H$_2$O masers in M33/19 and IC\,133 and
the proximate H\,I gas are in very good agreement ($<$ 10 km~s$^{-1}$). This
strongly suggests that the maser sources are rotating with the H\,I
gas in the galaxy. However, while the agreement between the rotation
model presented in \cite{24} and the radial velocity of the H\,I gas at the
position of IC\,133 is also very good ($<$ 5 km~s$^{-1}$), there is a
difference of $\sim$ 15 km~s$^{-1}$ at the position of M33/19. Hence,
we conservatively assume a systematic error of 20 km~s$^{-1}$ in each
velocity component for the relative velocity of the two maser
components. Comparing the measured angular motion of 30.8 $\pm$ 4
$\mu$as yr$^{-1}$ in right ascension with the expected linear motion
of 106 $\pm$ 20 km~s$^{-1}$, one gets a geometric distance of
\begin{eqnarray}
\nonumber
D=730 \pm 135 \pm 100~\mathrm{kpc},
\end{eqnarray}
where the first error indicates the systematic error from
the rotation model while the second error is the statistical error from the 
VLBI proper motion measurements.

After less than three years of observations, the uncertainty in the
distance estimate is already dominated by the uncertainty of the rotation
model of M33. However, this can be improved in the near future by
determining a better rotation model using higher resolution (e.g., Very Large
Array or Westerbork Synthesis Radio Telescope) data  of H\,I gas in the
inner parts of the disk. Also, the precision of the proper motion
measurements will increase with time as $t^{3/2}$ for evenly spaced
observations.

Within the current errors the geometric distance of $730 \pm 100 \pm
135~\mathrm{kpc}$ is in good agreement with recent Cepheid and Tip of
the Red Giant Branch (TRGB) distances of 802$\pm$51 kpc
\cite{25} and 794$\pm$23 kpc \cite{26} respectively.

\subsection{Proper Motion of M33}

The observed proper motion $\tilde{\vec v}_{prop}$ of a maser 
(M33/19 or IC\,133) in M33 can be decomposed into three components
$\tilde{\vec v}_{prop}=  \vec v_{rot} + \vec v_\odot + \vec v_{M33}$.
Here $\vec v_{rot}$ is the motion of the masers due to the internal
galactic rotation in M33 and  $\vec v_\odot$ is the apparent motion of
M33 caused by the rotation of the Sun around the Galactic Center. The last
contribution $\vec v_{M33}$ is the true proper motion of M33 relative
to the Galaxy. 

Since the motion of the Sun \cite{18,20} and the rotation of M33 \cite{24}
is known, one can calculate the two contributions $\vec v_{rot}$ and 
$\vec v_\odot$. Combining these velocity vectors, one gets the true proper 
motion of M33:

\begin{eqnarray}
\nonumber
\dot\alpha_{M33}=&\dot{\tilde\alpha}_{prop}-\dot\alpha_{rot}-\dot\alpha_\odot\\\nonumber
=&-29.3\pm7.6 \frac{\mathrm{\mu as}} {\mathrm{yr}}=-101\pm35 \frac{km}{s}\\\nonumber
\mathrm{and}\\\nonumber
\dot\delta_{M33}=&\dot{\tilde\delta}_{prop}-\dot\delta_{rot}-\dot\delta_\odot\\\nonumber
=&45.2\pm9.1 \frac{\mathrm{\mu as}} {\mathrm{yr}}=156\pm47 \frac{km}{s}.
\end{eqnarray}

Finally, the
systemic radial velocity of M33 is $-$179 km~s$^{-1}$ \cite{24}. The
radial component of the rotation of the Milky Way toward M33 is $-$140
$\pm$ 9 km~s$^{-1}$.  Hence, M33 is moving with $-$39 $\pm$ 9
km~s$^{-1}$ toward the Milky Way. This gives now the three dimensional
velocity vector of M33 which is plotted in Fig.~\ref{LG}. The
total velocity of M33 relative to the Milky Way is 190 $\pm$ 59
km~s$^{-1}$.

\subsection{Proper Motion of IC\,10}

\begin{figure}
\begin{center}
\includegraphics[width=8cm,angle=-90]{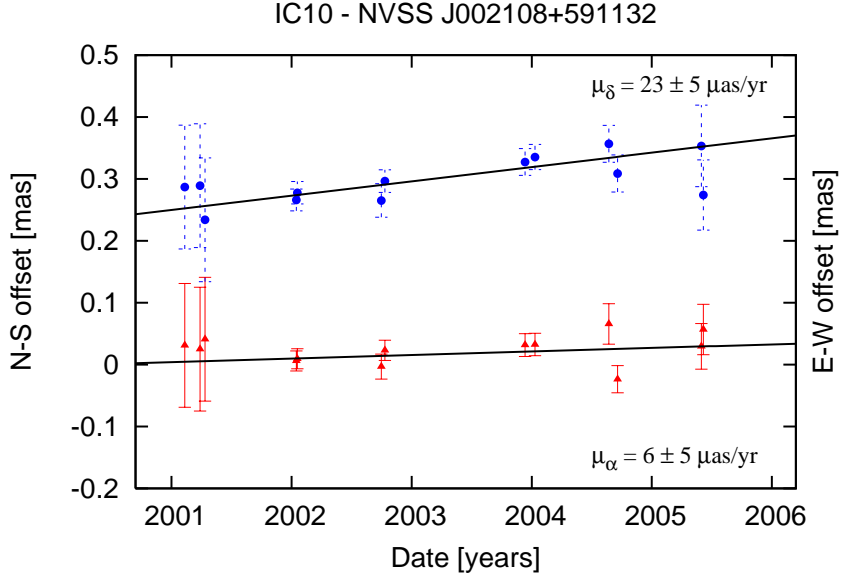}
\caption{The position of the maser in IC\,10 relative to a continuum background
source in East-West (red triangles) and North-South (blue circles). The lines 
show a variance weighted linear fit to the data points. Taken from 
\protect\cite{28}.}
\end{center}
\label{pos_ic10}
\end{figure}

In IC\,10 only the strongest maser component was detected in all epochs.
The position offsets
of this maser feature in IC\,10 are shown in Fig.~\ref{pos_ic10}.
The uncertainties in the observations of the first epoch are
larger than the others, because no geodetic-like observations were
done to compensate the zenith delay errors. A rectilinear motion was fit to
the data and yielded a value of 6$\pm$5
$\mu$as~yr$^{-1}$ toward the East and 23$\pm$5 $\mu$as~yr$^{-1}$ toward the
North.

Once again, the contributions $\vec v_{rot}$ and $\vec v_\odot$ can be 
calculated from the known motion of the Sun \cite{18,20} and the known 
rotation of IC\,10 \cite{29,30}. The true proper motion of IC\,10 is then 
given by:

\begin{eqnarray}
\nonumber\dot\alpha_{IC\,10}&=&\dot{\tilde\alpha}_{prop} - \dot\alpha_{rot} -\dot\alpha_\odot\\\nonumber
&=&-39\pm9~\mu\mathrm{as}~\mathrm{yr}^{-1}
=-122\pm31~\mathrm{km}~\mathrm{s}^{-1}\\\nonumber
\mathrm{and}\\\nonumber
\dot\delta_{IC\,10}&=&\dot{\tilde\delta}_{prop} - \dot\delta_{rot} -\dot\delta_\odot\\\nonumber
&=&31\pm8~\mu\mathrm{as}~\mathrm{yr}^{-1}=97\pm27~{\mathrm{km}}~\mathrm{s}^{-1}
\end{eqnarray}

\begin{figure}
\begin{center}
\includegraphics[bbllx=2.5cm,bburx=17.5cm,bblly=10.5cm,bbury=27cm,clip=,width=16cm,angle=0]{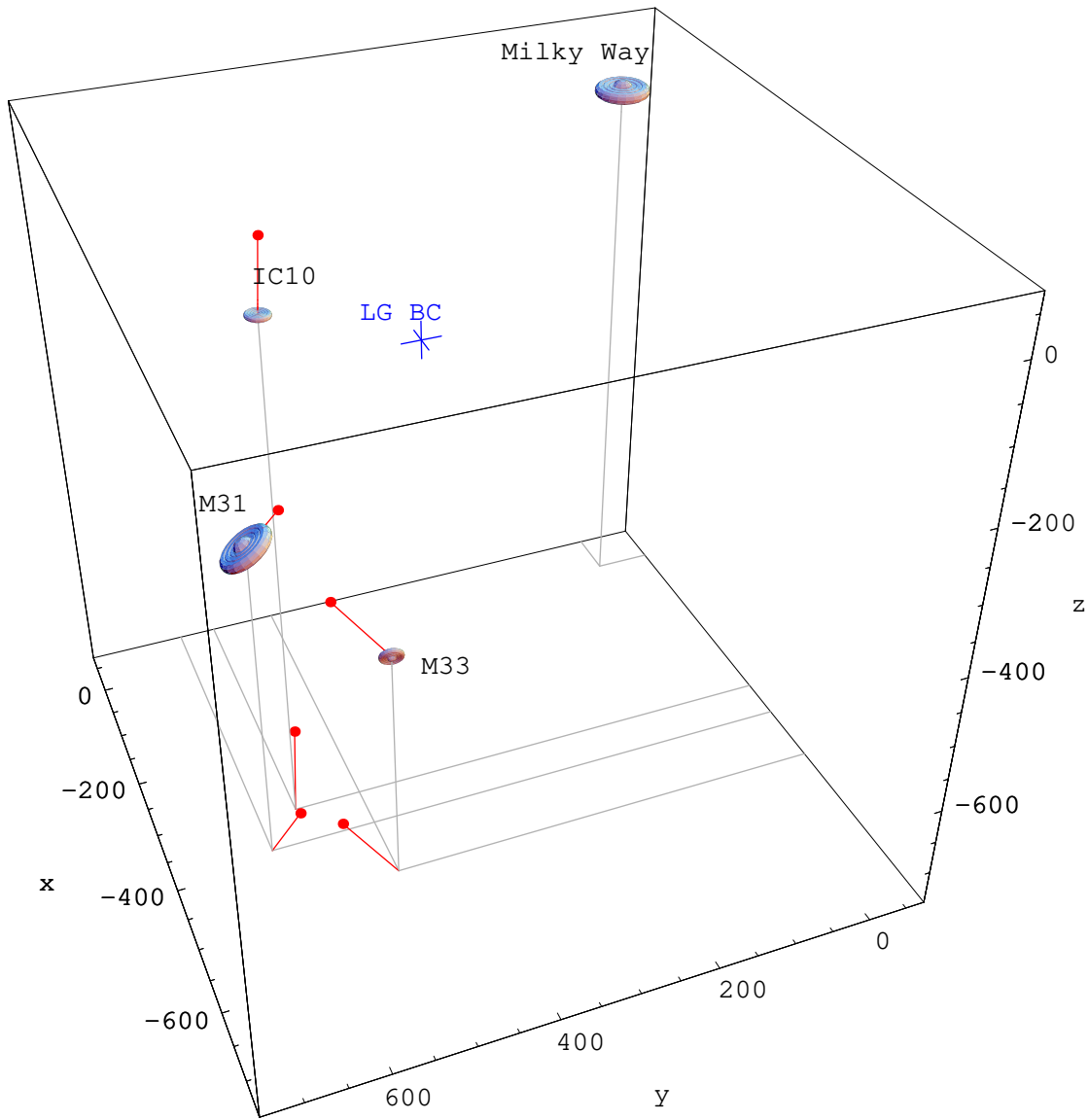}
\caption{Schematic view of the Local Group with the space velocity of
  M33 and the radial velocity of Andromeda. The blue cross marks the
  position of the Local Group Barycenter (LG BC) according to
  \protect\cite{27}. Taken from \protect\cite{28}.}
\label{LG}
\end{center}
\end{figure}

The measured systematic heliocentric velocity of IC\,10 ($-344\pm3$
km~s$^{-1}$,~\cite{31}) is the sum
of the radial motion of IC\,10 toward the Sun and the component of the
solar motion about the Galactic Center toward IC\,10 which is
-196$\pm$10 km~s$^{-1}$. Hence, IC\,10 is moving with 148$\pm$10
km~s$^{-1}$ toward the Sun.
The proper motion and the radial velocity combined give the
three-dimensional space velocity of IC\,10. This velocity vector is
shown in the schematic view of the Local Group in Fig.~\ref{LG}. The 
total velocity is 215$\pm$42~km~s$^{-1}$ relative to the Milky Way.

\section{Local Group Dynamics and Mass of M31}

If IC\,10 or M33 are bound to M31, then the velocity of the two galaxies
relative to M31 must be smaller than the escape velocity and one can
deduce a lower limit on the mass of M31:

\begin{eqnarray}
M_{M31}>\frac{v_{rel}^2R}{2G}.\nonumber
\end{eqnarray}

\begin{figure}
\center{M$_\mathrm{M31}$ [M$_\odot$]}
{\includegraphics[width=9cm,bbllx=0cm,bburx=35.5cm,bblly=1.0cm,bbury=5.5cm,clip=,angle=0]{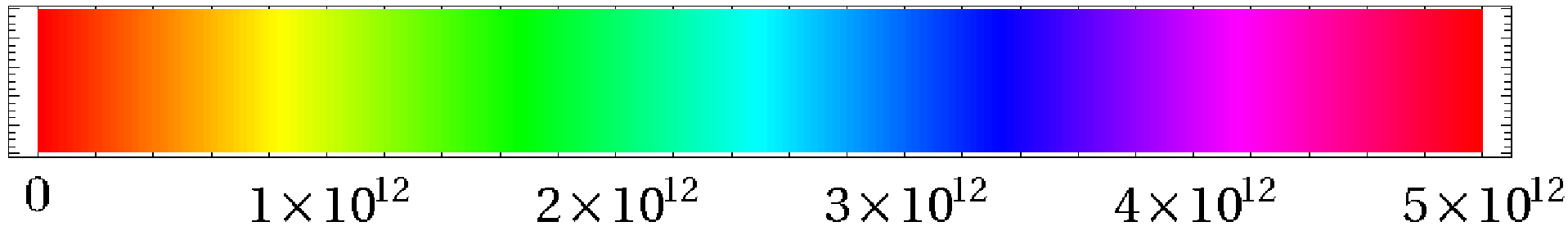}}
{\includegraphics[width=9cm,bbllx=2.5cm,bburx=13cm,bblly=16.9cm,bbury=25.5cm,clip=,angle=0]{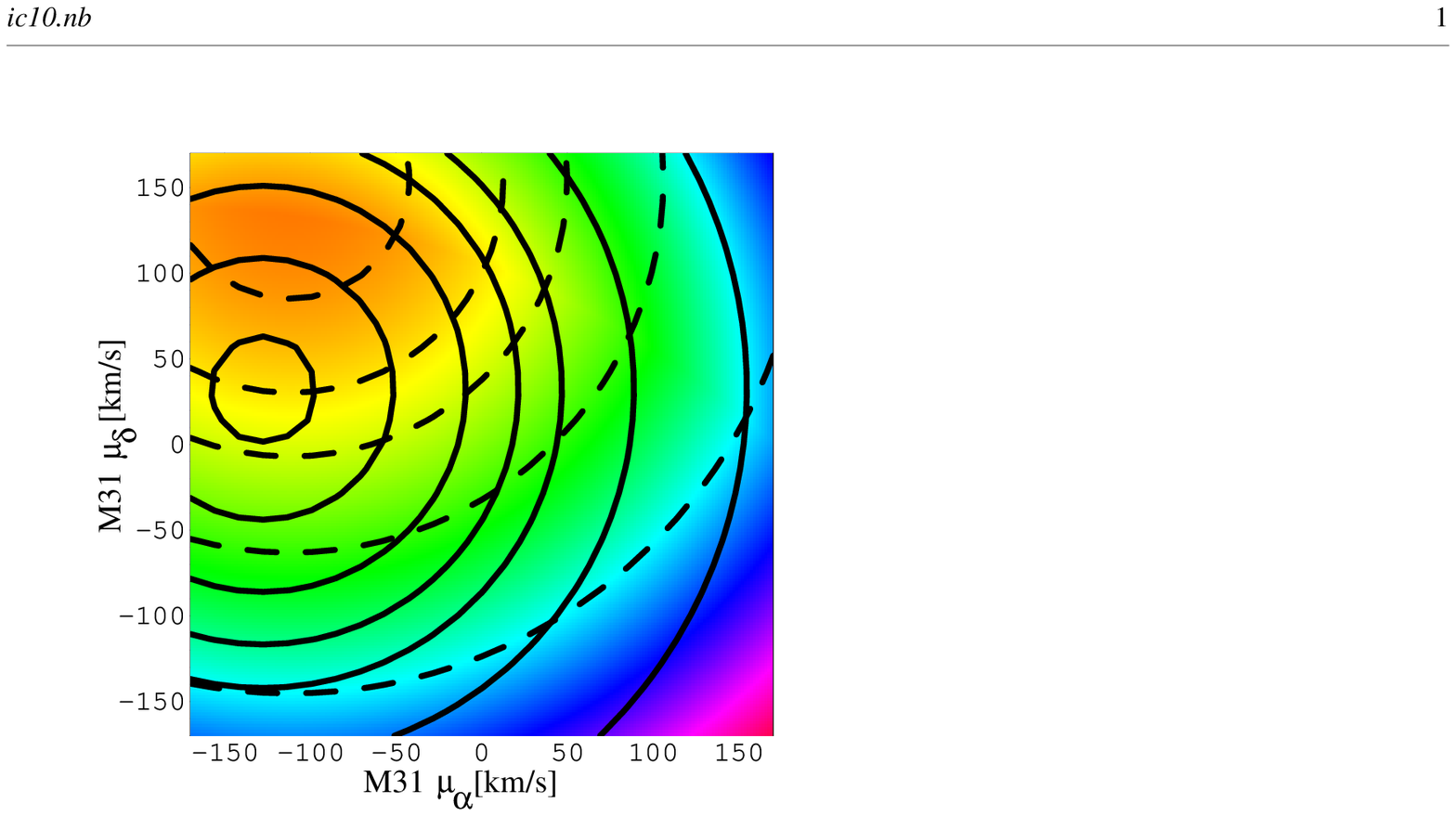}}
{\includegraphics[width=9cm,bbllx=2.5cm,bburx=13cm,bblly=16.9cm,bbury=25.5cm,clip=,angle=0]{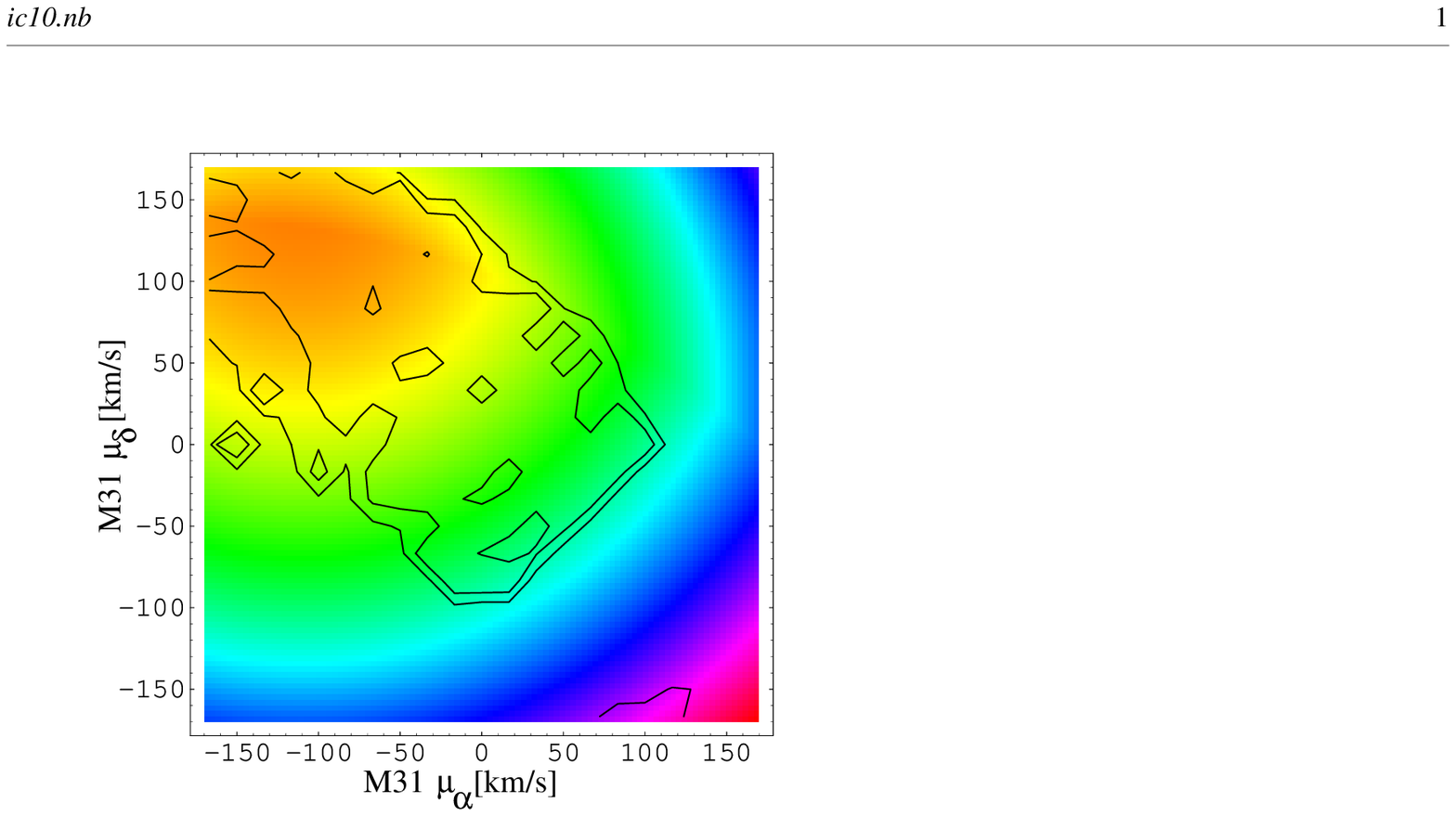}}
\caption{{\bf Top:} Lower limit on the mass of M31 for different tangential 
motions of M31 assuming that M33 (dashed) or IC\,10 (solid) are bound to M31. 
The lower limits are (4, 5, 7.5, 10, 15, 25)$\times10^{11}$M$_\odot$ for M33, 
and (0.7, 1, 2.5, 5, 7.5, 10, 15, 25)$\times10^{11}$M$_\odot$ for IC\,10, 
rising from inside. The colour scale indicates the maximum of both values.
{\bf Bottom:} The colour scale is the same as above and gives the
lower limit on the mass of M31. The contours show ranges of proper
motions that would have lead to a large amount of stars stripped from the disk
of M33 through interactions with M31 or the Milky Way in the past.
The contours delineate 20\% and 50\%  of the total number of stars stripped
\protect\cite{32}. These regions can be excluded, since the
stellar disk of M33 shows no signs of such interactions. Taken from 
\protect\cite{28}.}
\label{mass-m31}
\end{figure}

A relative velocity of 147 km~s$^{-1}$ -- for a zero tangential motion of M31
-- and a distance of 262 kpc between IC\,10 and M31 gives a lower limit of
6.6 $\times 10^{11}$M$_\odot$. One can repeat this calculation for any
tangential motion of M31. The results are shown in Fig.~\ref{mass-m31} (top).
The lowest
value of 0.7 $\times 10^{11}$M$_\odot$ is found for a tangential motion of M31
of --130 km~s$^{-1}$ toward the East and 35 km~s$^{-1}$ toward the North.

For a relative motion of 230 km~s$^{-1}$ between M33 and M31 -- again for a
zero tangential motion of M31 -- and a distance of 202 kpc, one gets a lower
limit of 1.2 $\times 10^{12}$M$_\odot$ \cite{22}.
Fig.~\ref{mass-m31} (top) shows also the lower limit of the mass of M31 for
different
tangential motions of M31 if M33 is bound to M31. The lowest value is
4 $\times 10^{11}$M$_\odot$ for a tangential motion of M31 of
--115 km~s$^{-1}$ toward the East and 160 km~s$^{-1}$ toward the North.

In \cite{32} it is
found that proper motions of M31 in negative right ascension and positive
declination would have lead to close interactions between M31 and M33 in the
past. These proper motions of M31 can be ruled out, since the stellar disk of
M33 does not show any signs of strong interactions.

Thus, we can rule out certain
regions in Fig.~\ref{mass-m31}. This results in a lower limit of
7.5$\times 10^{11}$M$_\odot$ for M31 and agrees with a recent
estimate of $12.3^{+18}_{-6}\times10^{11}$~M$_\odot$ derived from the
three-dimensional positions and radial velocities of its satellite
galaxies \cite{33}.

\section{Summary}
More than 80 years after van Maanen's observation, we have succeeded
in measuring the rotation and proper motion of M33 as well as the proper
motion of IC\,10. These measurements provide a new handle on dynamical 
models for the Local Group and the mass and dark matter halo of Andromeda 
and the Milky Way.

We have presented astrometric VLBA observations of the H$_2$O masers in the
Local Group galaxies M33 and IC\,10. We measured the proper motion of the
masers relative to background quasars. Correcting for the internal rotation
of M33, IC\,10 and the rotation of the Milky Way these measurements yield  
proper motions of the two galaxies. The total space velocities relative to the
Milky Way of M33 and IC\,10 are 190$\pm$59 km~s$^{-1}$ and 215$\pm$42 
km~s$^{-1}$, respectively.
If IC\,10 and M33 are bound to M31, one can calculate a lower limit
of the mass of M31 of 7.5~$\times 10^{11}$M$_\odot$.

Further VLBI observations within the next few years and an improved
rotation model have the potential to improve the accuracy of the
distance estimate to less than 10$\%$. At least one additional maser
source exists in M33 \cite{21} that will be used in the future to increase 
the accuracy of the measurements. A third region of maser activity will 
also help to check for non-circular velocities of the masers.

Unfortunately, no maser emission could be found in Andromeda despite
intensive searches. Hence, the
proper motion vector of Andromeda is still unknown.
In the near future, new technical developments using higher bandwidths
will increase the sensitivity of VLBI. This will allow the detection of
radio emission from the central black hole of Andromeda, M31*, with
flux densities of $\sim$30 $\mu$Jy \cite{34}, to
measure the proper motion of Andromeda.

Today we are only able to study the extreme (bright) tip of the maser
luminosity distribution for interstellar masers. The Square Kilometer
Array (SKA), with substantial collecting area on intercontinental
baselines and a frequency coverage up to 22 GHz
\cite{35}, will provide the necessary sensitivity to
detect and measure the proper motions of a much greater number of
masers in active star forming regions in the Local Group.

\end{document}